\documentclass[aip,superscriptaddress,showpacs,nofootinbib,floatfix,reprint]{revtex4-1}
\usepackage{amsmath}
\usepackage{amssymb}
\usepackage{graphicx}
\usepackage{float}
\usepackage{subfigure}
\usepackage{color}
\usepackage[utf8x]{inputenc}

\begin{document}
\title{Title}
\author{Nathann T. Rodrigues}
\email{nathan.rodrigues@ufv.br}
\author{Tiago J. Oliveira}
\email{tiago@ufv.br}
\affiliation{Departamento de Física, Universidade Federal de Viçosa, 36570-900, Vi\c cosa, MG, Brazil}
\date{\today}

\title{Three stable phases and thermodynamic anomaly in a binary mixture of hard particles}

\begin{abstract}
While the realistically modeling of the thermodynamic behavior of fluids usually demands elaborated atomistic models, much have been learned from simplified ones. Here, we investigate a model where point-like particles (with activity $z_0$) are mixed with molecules that exclude their first and second neighbors (i.e., cubes of lateral size $\lambda=\sqrt{3}a$, with activity $z_2$), both placed on the sites of a simple cubic lattice with parameter $a$. Only hard-core interactions exist among the particles, so that the model is athermal. Despite its simplicity, the grand-canonical solution of this model on a Husimi lattice built with cubes revels a fluid-fluid demixing, yielding a phase diagram with two fluid phases (one of them dominated by small particles - $F0$) and a solid-like phase coexisting at a triple-point. Moreover, the fluid-fluid coexistence line ends at a critical point. An anomaly in the total density ($\rho_T$) of particles is also found, which is hallmarked by minima in the isobaric curves of $\rho_T$ versus $z_0$ (or $z_2$). Interestingly, the line of minimum density cross the phase diagram starting inside the region where both fluid phases are stable, passing through the $F0$ one and ending deep inside its metastable region, in a point where the spinodals of both fluid phases cross each other.
\end{abstract}

\maketitle

\section{Introduction}
\label{secIntro}

Lattice gas (LG) models have a long history in the atomistic modeling of fluids, among a wide variety of other systems \cite{BarrySimon}. In general, two key ingredients are expected to be necessary to reproduce the gas-solid-liquid phase behavior of simple fluids: repulsive excluded volume interactions, which can naturally be imposed by the particle size and lattice structure in LG systems; and short-range attractive interactions. The first one alone, even in the simplest hard-core (athermal) case, is capable to yield entropy-driven fluid-solid transitions, provided that the range of exclusion (associated with the lateral particle size $\lambda$ and its placement on the lattice) is larger than the lattice spacing $a$. The ordering transition in these hard-LGs is usually studied considering molecules which exclude up to their $k$th nearest-neighbors (NN) - the so-called $k$NN models. These are discrete versions of hard disks and hard spheres in two- and three-dimensions, respectively, which have been numerically investigated for a broad range of $k$'s on the square (see \cite{Heitor,Rajesh} and refs. therein) and simple cubic lattices \cite{Panagiotopoulos}. The 1NN case on the triangular lattice is the famous hard hexagon model exactly solved by Baxter \cite{baxterHH,baxterBook}. Entropy-driven transitions in hard-LGs with particles of several other shapes have also been considered, such as triangles \cite{Nienhuistri}, dimers \cite{dimers}, rectangles \cite{rectangles}, rods \cite{rods}, Y-shaped \cite{RajeshY}, cubes \cite{Rajeshcubes}, etc.

In mixtures of hard particles the situation becomes much more interesting, once demixing transitions may also take place in the system. For instance, in binary nonadditive mixtures of hard particles - whose minimum distance $\sigma_{ij}$ between particles $i$ and $j$ follows $\sigma_{AB}>(\sigma_{AA}+\sigma_{BB})/2$ - a demixing is expected since the space is filled more effectively by the pure phases than by the mixture. Fluid-fluid demixing has been indeed observed in different theoretical \cite{Widom,Melnyk,Santos,Louis} and numerical approaches \cite{Dijkstra,Schmidt,Brader,frenkel0nn1nn,Dickman95} for binary nonadditive mixtures of isotropic hard particles. Mixtures of molecules with anisotropic shapes are also known to undergo such transition \cite{Roiji,Wensink,Dubois,Varga,Mederos,Schmidt02,Heras}. For additive mixtures - where $\sigma_{AB} = (\sigma_{AA}+\sigma_{BB})/2$ - the situation is more controversial: while the possibility of fluid-fluid phase separation has been raised in some analytical studies of hard-spheres, provided that the particle's sizes are dissimilar enough \cite{BH91,Lekkerkerker932}, it seems that the demixing is usually preempted by a fluid-solid or solid-solid coexistence [see, e.g, Refs. \cite{LafuenteCuesta,LafuenteCuesta2} for a detailed discussion and references]. 

Once $k$NN systems can display fluid-solid transitions and since binary mixtures of them are nonadditive, we are immediately led to inquiry whether such mixtures display fluid-fluid demixing, yielding phase diagrams with three stable phases. This important point was firstly addressed by Frenkel and Louis \cite{frenkel0nn1nn}, investigating a mixture of hard hexagons (1NN) with point particles (0NN) on the triangular lattice, but no fluid-fluid transition was found. Subsequently, Van Duijneveldt and Lekkerkerker \cite{Lekkerkerker93,Lekkerkerker95} claimed to have found three stable (gas-liquid-solid) phases for this system. However, their results were contested by Verberkmoes and Nienhuis (VN) in Ref. \cite{Nienhuis}, where strong numerical evidence were presented that such mixture has in fact only the fluid and solid phases reminiscent from the simple hard hexagon model. The mixture of point particles with 1NN ones has been also studied by different methods on the square lattice \cite{poland,Jim01,tiago15}. The phase diagram of this system presents again only a fluid and a solid phase, which are separated by a critical and a coexistence line meeting at a tricritical point. The same properties were also found in an exact (mean-field) solution of this model on the Bethe lattice \cite{tiago11}. It is noteworthy that the same scenario was suggested for the hard hexagons with point particles by VN \cite{Nienhuis}. In three dimensions, as far as we known, there exits no study of mixtures of $k$NN molecules. Therefore, it seems that the existence of three stable (``gas-liquid-solid'') phases arising in LG systems consisting of binary mixtures of $k$NN particles is still an open issue.

Here, we investigate the (nonadditive) mixture of 0NN with 2NN molecules placed on the simple cubic lattice and demonstrate that beyond the solid phase, two stable fluid phases are also present in the phase diagram. Actually, we present a mean-field approximation for the model on the cubic lattice, consisting of its semi-analytical grand-canonical solution on a Husimi lattice - the core of a Cayley tree - built with cubes (see Fig. \ref{fig1}a). As an aside, we remark that this kind of solution on hierarchical lattices advance over other typical mean-field treatments because some correlations are still present in the system \cite{Gujrati}. For this reason, they not only usually provide the qualitatively correct thermodynamic behavior for the analyzed models (as is the case, for instance, in the 0NN-1NN mixture on the square and Bethe lattices \cite{tiago11,tiago15}), but for some LG models they yield even some quantitative agreement with results from Monte Carlo simulations on regular lattices \cite{Buzano,tiago10}. In some LG systems (e.g, with nematic order), however, more elaborated lattices may be need \cite{RajeshJurgen1}. In our present study, beyond the fluid and solid phases expected for the simple 2NN model \cite{Panagiotopoulos}, a second disordered (fluid-like) phase is also stable for large activity of 0NN particles, which is featured by a dominance of such particles. A rich phase diagram is obtained, with this 0NN fluid ($F0$) phase coexisting with the regular fluid ($RF$) and solid ($S$) ones at a triple point, where the $RF$-$S$, $RF$-$F0$ and $F0$-$S$ coexistence lines meet. The fluid-fluid coexistence line ends at a critical point. A density anomaly, characterized by minima in the isobaric density curves, is observed in this system, indicating that it can be very useful as a starting point for understanding the behavior of more complex fluids. 

The rest of the paper is organized as follows. In Sec. \ref{secModel} we define the model and solve it on the Husimi lattice in terms of recursion relations. The thermodynamic properties of the model are presented in Sec. \ref{secResults}. Our final discussions and conclusions are summarized in Sec. \ref{secConc}. The calculation of the free energy is devised in the appendix.

\section{Model definition and solution in terms of recursion relations}
\label{secModel}

\begin{figure}[!t]
\includegraphics[width=8.5cm]{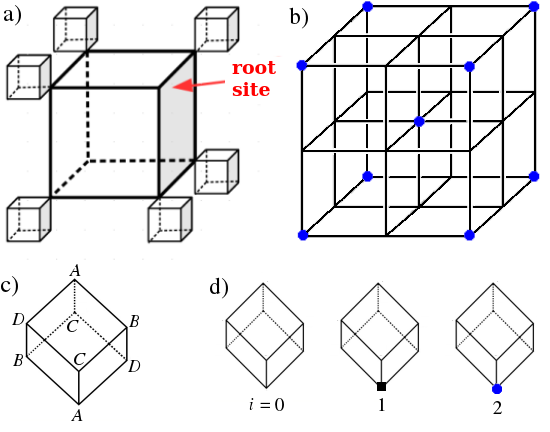}
\caption{(a) Illustration of part of a Husimi lattice built with cubes. (b) The ground state of the solid phase (the limit $z_2 \rightarrow \infty$, with $z_0=0$) in the cubic lattice. (c) Definition of sublattices in an elementary cube. (d) Possible states of the root site, where 0NN and 2NN particles are indicated by a square and a circle, respectively.}
\label{fig1}
\end{figure}

We consider a lattice gas with a binary mixture of hard particles placed on (and centered at) the vertices of a cubic lattice. Assuming that the lattice spacing is $a$, the small particles (0NN) are cubes of lateral size $\lambda=a$, with faces parallel to the lattice, so that they occupy a single lattice site and do not exclude their neighbors. On the other hand, the large particles (2NN) are cubes of lateral size $\lambda = \sqrt{3}a$ slanted in a way that they exclude their first and second nearest neighbors. Activities $z_0$ and $z_2$ are associated with 0NN and 2NN particles, respectively. Thereby, for $z_0=0$ one recovers the simple 2NN model, which have already been investigated through Monte Carlo simulations on the cubic lattice and displays a discontinuous fluid-solid transition \cite{Panagiotopoulos}. Moreover, in the limit $z_0 \rightarrow \infty$, with $z_2$ finite, all sites are occupied with 0NN particles, leading to the densities $\rho_0 = 1$ and $\rho_2=0$ for the densities of small and large particles, respectively. In the opposite limit ($z_2 \rightarrow \infty$, with $z_0$ finite), however, only $1/4$ of the lattice sites can be occupied by the large particles, so that $\rho_2 = 1/4$ and $\rho_0 = 0$. In this case, one has the ground state of the solid phase, which is featured by a sublattice ordering, as shows Fig. \ref{fig1}b. Namely, dividing the lattice into four sublattices ($A$, $B$, $C$ and $D$), composed by the two diametrically opposite sites in each elementary lattice cube (see Fig. \ref{fig1}c), the solid phase is hallmarked by a symmetry breaking, such that one of these sublattices is more populated than the others.

In order to solve this model on a Husimi lattice built with cubes, we define partial partition functions (ppf's) according to the state of the root site of an elementary cube. As shows Fig. \ref{fig1}d, for each sublattice there are three possible states: the root site can be empty ($i=0$), occupied by a 0NN ($i=1$) or a 2NN particle ($i=2$). Since there are four sublattices, one has a total of twelve possible states for the root site. For sake of simplicity, for a while let us consider that it is in sublattice $A$. Note that by attaching seven cubes to the vertices of another cube, with exception of the root site, we obtain a subtree with 1 generation. Then, by attaching the root sites of seven of these subtrees to the vertices of another new cube, a 2-generation subtree is built. Proceeding in this way, we can obtain a ($M+1$)-generation subtree from seven ones with $M$ generations. Now, by taking into account the allowed configurations of the model in the new cube, we can determine all the possible ways of attaching seven subtrees to it, keeping its root site in state $A_i$. This give us the ppf $A'_i$ in generation $M+1$ as a function of the ppf's $A_j$, $B_j$, $C_j$ and $D_j$ (with $j=0$, 1 and 2) in generation $M$. This simple procedure allow us to construct a set of recursion relations for the ppf's of the model, which for sublattice $A$ reads:

\begin{widetext}
\begin{subequations}
\begin{eqnarray}
A'_{0} &=& z_2 A_2 B_0^2 C_0^2 D_0^2+A_0 \left\lbrace 2 z_0 z_2 B_1 B_2 C_0^2 D_0^2 + z_2^2 B_2^2 C_0^2 D_0^2+z_0^2 B_1^2 (C_0+z_0 C_1)^2 (D_0+z_0 D_1)^2 \right. \\ \nonumber &+& \left. 2 B_0 \left[z_2 B_2 C_0^2 D_0^2 + z_0 B_1 (C_0+z_0 C_1)^2 (D_0+z_0 D_1)^2\right] + B_0^2 \left[ 2 z_0 z_2 C_1 C_2 D_0^2+ z_2^2 C_2^2 D_0^2+ z_0^2 C_1^2 (D_0+z_0 D_1)^2 \right.\right. \\ \nonumber &+& \left.\left. 2 C_0 (z_2 C_2 D_0^2+z_0 C_1 (D_0+z_0 D_1)^2) + C_0^2 (D_0+z_0 D_1+z_2 D_2)^2 \right] \right\rbrace + z_0 A_1 (B_0+z_0 B_1)^2 (C_0+z_0 C_1)^2 (D_0+z_0 D_1)^2
\end{eqnarray}

\begin{eqnarray}
A'_{1} = (A_0 + z_0 A_1) (B_0 + z_0 B_1)^2 (C_0 + z_0 C_1)^2 (D_0 + z_0 D_1)^2 + z_2 A_2 B_0^2 C_0^2 D_0^2
\end{eqnarray}

\begin{eqnarray}
A'_{2} &=& (A_{0} +  z_0 A_{1} +  z_2 A_{2}) B_{0}^{2} C_{0}^{2} D_{0}^{2} 
\end{eqnarray}
\label{eqRRs}
\end{subequations}
\end{widetext}

The ppf's for the other sublattices can be obtained from these ones by cyclic permutations of the letters ($A \rightarrow B$, $B \rightarrow C$, $C \rightarrow D$, and $D \rightarrow A$). Clearly, these recursion relations (RRs) shall diverge if they are iterated too many times, what corresponds to the building up of infinite subtrees, which is the desired thermodynamic limit. Thereby, we analyze ratios of them, defined as $R_1^{(A)}=A_1/A_0$ and $R_2^{(A)}=A_2/A_0$ for sublattice $A$, which remains finite in the thermodynamic limit. For the other sublattices the definition is the same, with $A$ replaced by $B$, $C$ or $D$. Therefore, from the twelve RRs for the ppf's we obtain eight RRs for these ratios.

The real and positive solutions (fixed points) of these RRs correspond to the thermodynamic phases of the model on the Husimi lattice. To determine the stability limits of each phase we calculate the Jacobian matrix, whose 64 entries are given by the derivatives $\partial {R'_{i}}^{(K)}/\partial R_{j}^{(W)}$, with $i,j=1$, 2 and $K,W=A$, $B$, $C$, $D$, calculated in a given fixed point. In regions of the parameter space ($z_0,z_2$) where the largest eigenvalue ($\Lambda$) of this matrix is smaller than one ($\Lambda<1$), the fixed point is stable, as well as the corresponding thermodynamic phase. The condition $\Lambda = 1$ defines the stability limits.   

The partition function ($Y$) can be obtained, similarly to the ppf's, by considering the operation of attaching the root sites of eight subtrees to the eight vertices of a central cube. It can be written in a compact form, e.g, as 
\begin{equation}
Y= A_{0} A'_{0} + z_0 A_{1} A'_{1} + z_2 A_{2} A'_{2} = (A_0 B_0 C_0 D_0)^2 y,
\label{eqY}
\end{equation}
Using Eqs. \ref{eqRRs} to write the expanded expression for $Y$, the densities of small and large particles at the central cube, in sublattice $K$, are given respectively by 
\begin{equation}
\rho_0^{(K)} = \frac{R_1^{(K)}}{8 Y} \frac{\partial Y}{\partial R_1^{(K)}} \quad \text{and} \quad \rho_2^{(K)} = \frac{R_2^{(K)}}{8 Y} \frac{\partial Y}{\partial R_2^{(K)}}.
\end{equation}
Thence, $\rho_j = \rho_j^{(A)} + \rho_j^{(B)} + \rho_j^{(C)} + \rho_j^{(D)}$ gives us the total density of small ($j=0$) and large ($j=2$) particles at the central cube.

Following the ansatz proposed by Gujrati \cite{Gujrati}, and discussed in detail in the appendix, the bulk free energy per site of each phase of the model can be calculated from
\begin{equation}
 \phi_b = -\frac{1}{8} \ln \left[ \frac{\left( R_{0}^{(A)} R_{0}^{(B)} R_{0}^{(C)} R_{0}^{(D)}\right)^2}{y^{6}} \right],
 \label{eqFE}
\end{equation}
where $R_0^{(A)}\equiv A'_0/\left(A_0 B_0^2 C_0^2 D_0^2\right)$ and the others are obtained from cyclic permutations of the labels ($A$, $B$, $C$, $D$). In regions where two or more phases are stable, the points (or lines) where the free energies of these phases are equal define the coexistence loci. We notice also that since each lattice site occupies a volume $v_0=a^3$, the pressure in our grand-canonical formalism is given by $P=-\phi_b/a^3$.

\section{Thermodynamic behavior of the model}
\label{secResults}

\subsection{2NN model}

Let us start discussing the simple 2NN model. Since $z_0=0$ in this case, one has $R_1^{(A)}=R_1^{(B)}=R_1^{(C)}=R_1^{(D)}=0$ and the recursion relations (RRs) considerably simplify. Even then, we are not able to calculate simple analytical expressions for their fixed points and, so, we  numerically estimate them by iterating the RRs. For small values of $z_2$ one finds a homogeneous solution $R_2^{(A)}= R_2^{(B)}=R_2^{(C)}=R_2^{(D)}$, which corresponds to a disordered fluid (F) phase where all sublattices are equally populated. This phase is stable for $z_2 \leqslant 15.1364$. For large $z_2$, on the other hand, we find four equivalent fixed points featured by a symmetry breaking, such that one of the four sublattices dominates. For example, if $A$ is the dominant sublattice, one has $R_2^{(A)} \gg R_2^{(B)}=R_2^{(C)}=R_2^{(D)}$. This corresponds to the ordered solid (S) phase, which is stable for $z_2 \geqslant 5.0620$.

Therefore, both phases coexist in the region $5.0620 \leqslant z_2 \leqslant 15.1364$ and, hence, the fluid-solid transition is discontinuous. The free energy of both phases are equal [$\phi_b^{(F)} = \phi_b^{(S)}$] at $z_{2} = 5.7932$, which defines the transition point. The particle densities at this point for the fluid and solid phases are given, respectively, by $\rho_2^{(F)} = 0.1551$ and $\rho_2^{(S)} = 0.2025$, confirming the discontinuous nature of the transition. This result is consistent with Monte Carlo simulations of this model on the cubic lattice, where a first-order transition was also found \cite{Panagiotopoulos}. In this case, notwithstanding, the transition point is located at $z_2 \approx 1.70$\footnote{Note that the chemical potential at coexistence ($\beta \mu = 2.18\pm 0.01$) reported in \cite{Panagiotopoulos} is related with the activity through $z_2 = e^{\beta \mu}/\sigma^3$, where $\sigma=\lambda/a=\sqrt{3}$ for the 2NN particles.}, which is considerably smaller than our value. In agreement with this, the densities at coexistence $\rho^{(F)} \approx 0.10$ and $\rho^{(S)} \approx 0.13$ estimated in~\cite{Panagiotopoulos} are also smaller than the ones found here. We remark that while mean-field approximations usually predicts smaller critical points for continuous transitions than their correct values, this is not necessarily a rule in discontinuous transitions. In our specific case, the tree-like structure of the Husimi lattice turns the ordering of 2NN particles more difficult than on the cubic lattice, demanding thus higher densities (and so a large $z_2$).

\subsection{0NN-2NN model}

Now, we turn to the analysis of the properties of the full model, with two kinds of particles. For small $z_0 >0$ one still find the fluid and solid ($S$) phases reminiscent from the 2NN model, but now $R_1^{(K)} \neq 0$, for $K=A$, $B$, $C$ and $D$. The fluid phase is still homogeneous, so that $R_i^{(A)}=R_i^{(B)}=R_i^{(C)}=R_i^{(D)}\equiv R_i$, while in the solid phase there still exists a symmetry breaking and one of the sublattices dominates [e.g, $R_i^{(A)} \approx 1$ and $R_i^{(B)}=R_i^{(C)}=R_i^{(D)} \approx 0$, for $i=1,2$]. Beyond these expected phases, a remarkable result found here is the existence of a second homogeneous fluid phase, which is stable in the system for large values of $z_0$. In such phase, one also has $R_i^{(A)}=R_i^{(B)}=R_i^{(C)}=R_i^{(D)}\equiv R_i$, but with $R_1 \approx 1$ and $R_2 \approx 0$. Namely, this phase is featured by a dominance of the RRs associated with small particles and, so, hereafter we will refer to it as the 0NN fluid ($F0$) phase. In contrast, in the regular fluid ($RF$) phase the values of $R_1$ and $R_2$ depend more on the activities ($z_0$ and $z_2$).

\begin{figure}[!t]
\includegraphics[width=8.cm]{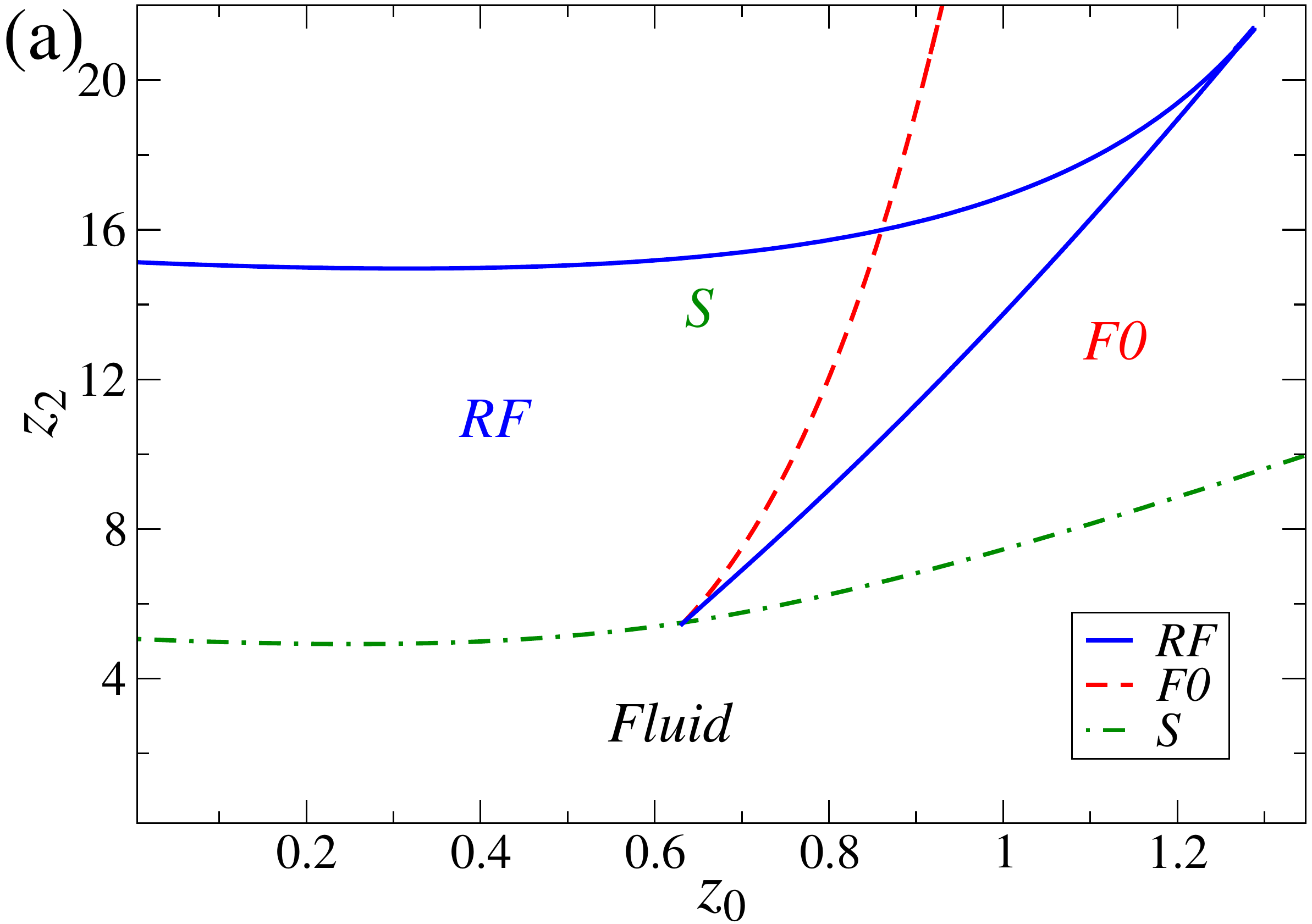}
\includegraphics[width=8.cm]{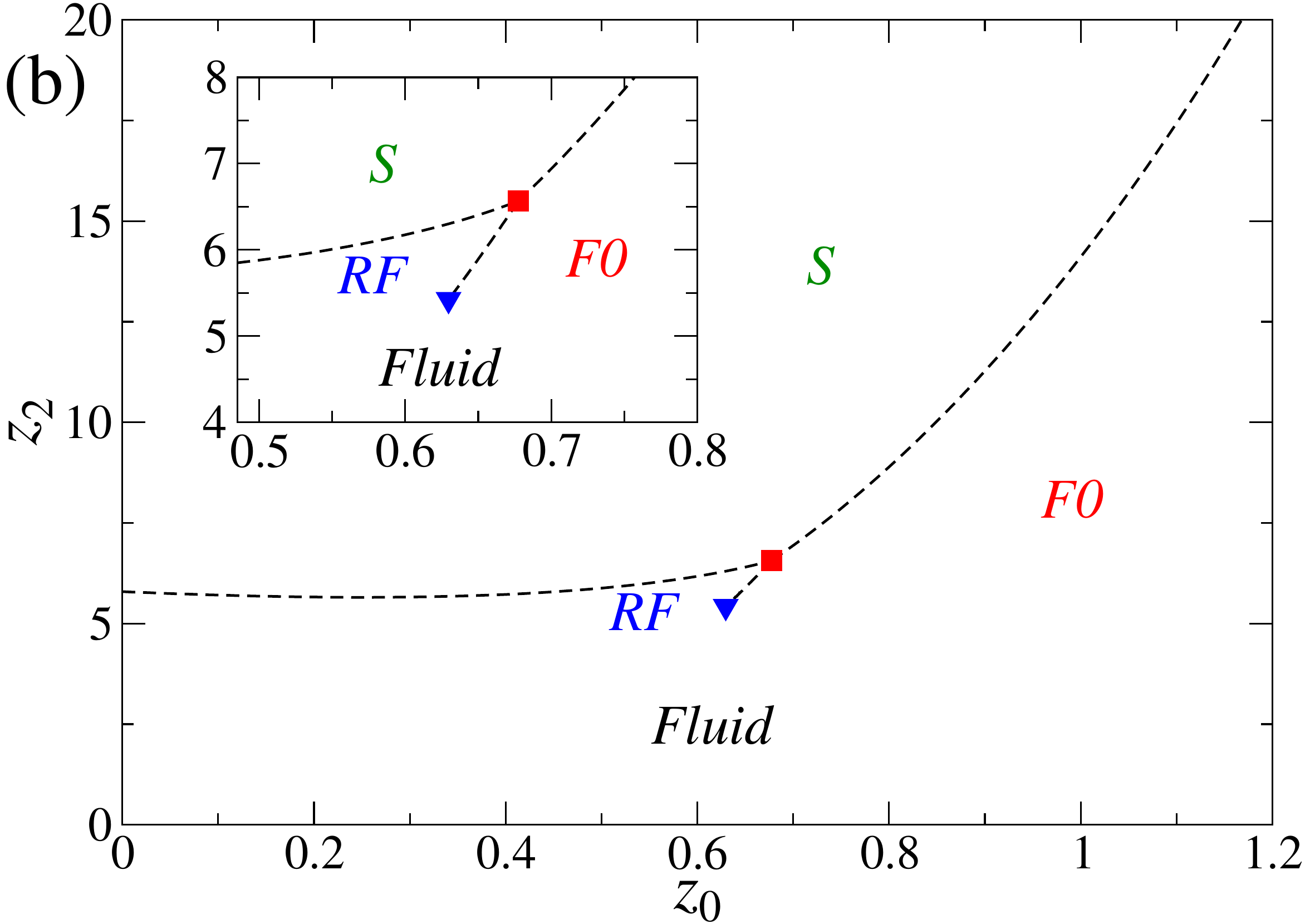}
\includegraphics[width=8.cm]{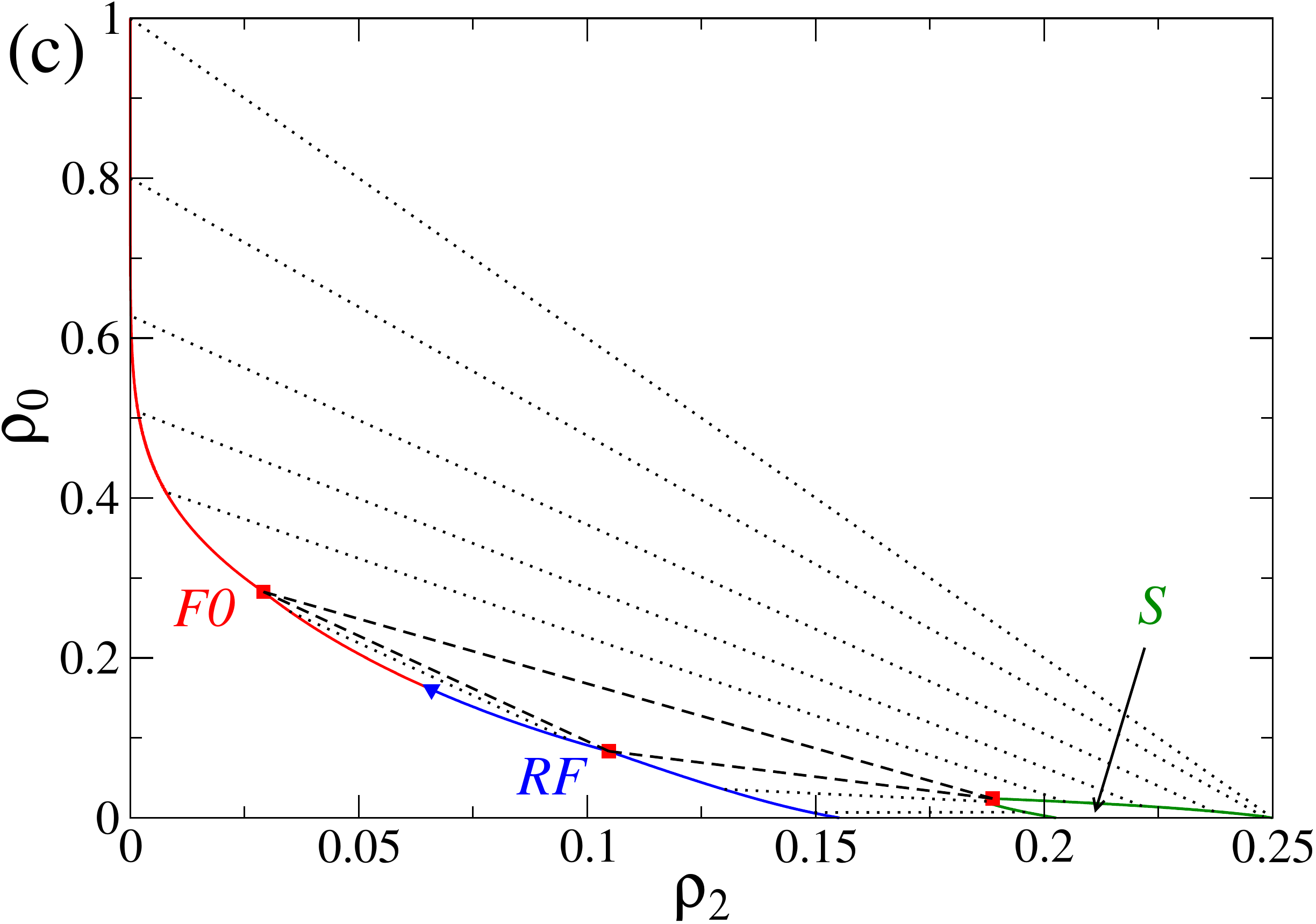}
\caption{(a) Stability limits of the regular fluid ($RF$), 0NN fluid ($F0$) and solid ($S$) phases. For small $z_2$, the $RF$ and $F0$ phases become indistinguishable in a single \textit{Fluid}. (b) Phase diagram in space ($z_0,z_2$). The dashed lines are the coexistence loci, while the square (red) and triangle (blue) symbols are the triple and critical points, respectively. The inset highlights the region around these points. (c) Phase diagram in density ($\rho_2,\rho_0$) space. The solid, dashed and dotted lines are the coexistence loci, the triple point and tie lines, respectively.}
\label{fig2}
\end{figure}

Figure \ref{fig2}a presents the stability limits (the spinodals) of these phases in the ($z_0,z_2$) space. We find that the $S$ phase is stable for any value of $z_0$ provided that $z_2$ is large enough. Conversely, the $F0$ phase is stable for any value of $z_2$ for large $z_0$'s. The $RF$ phase, on the other hand, is stable only in a limited region of the parameter space where $z_0$ and $z_2$ are both not too large. Interestingly, the spinodals of the $RF$ and $F0$ phases meet each other, and end, at the point $(z_{0,c},z_{2,c}) = (0.6297,5.4243)$, which turns out to be a critical point (CP). Hence, in the region below this point (for $z_2 < z_{2,c}$) we have a single fluid, since we cannot distinguish between the $RF$ and $F0$ phases.

\begin{figure}[!t]
\includegraphics[width=8.cm]{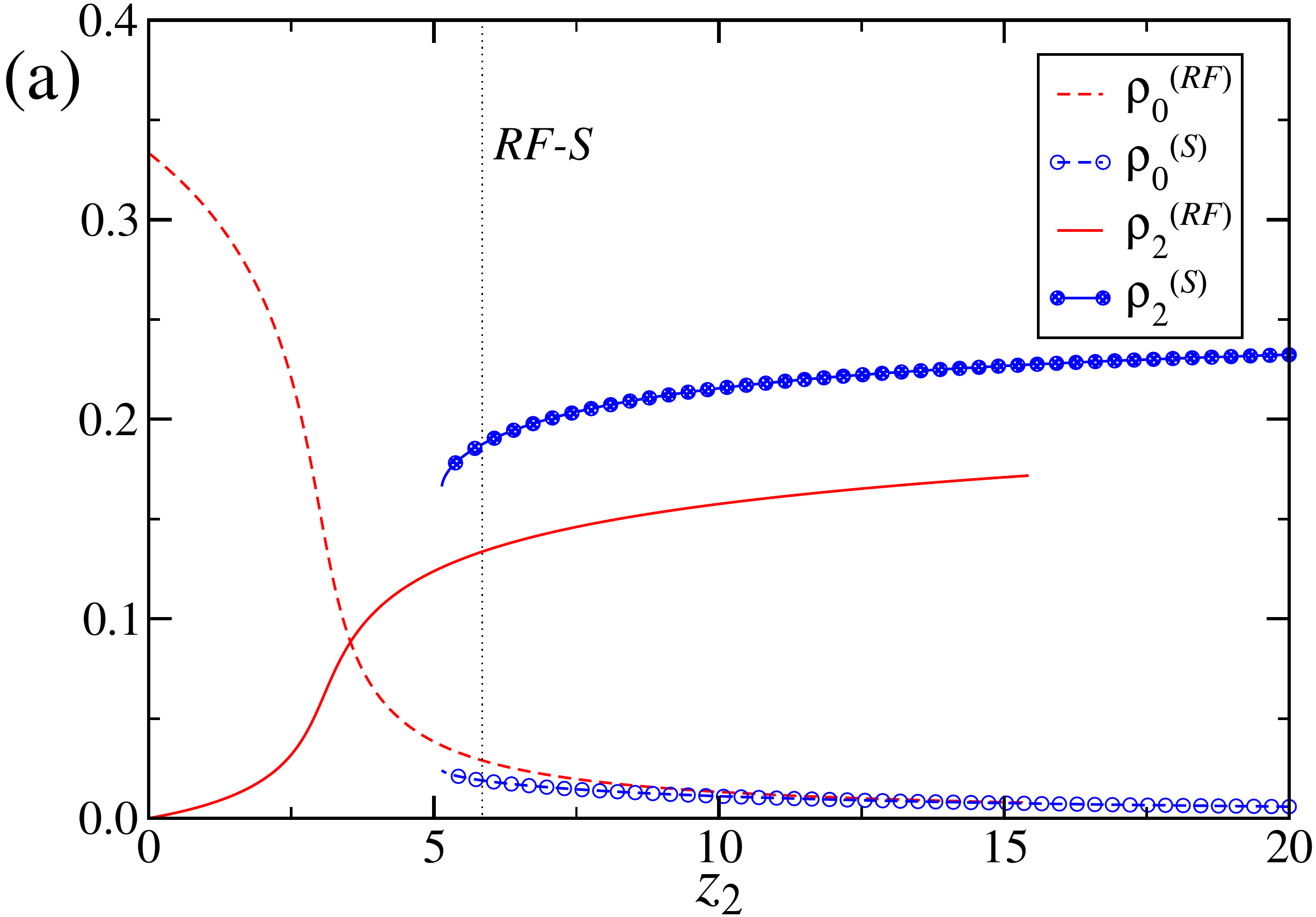}
\includegraphics[width=8.cm]{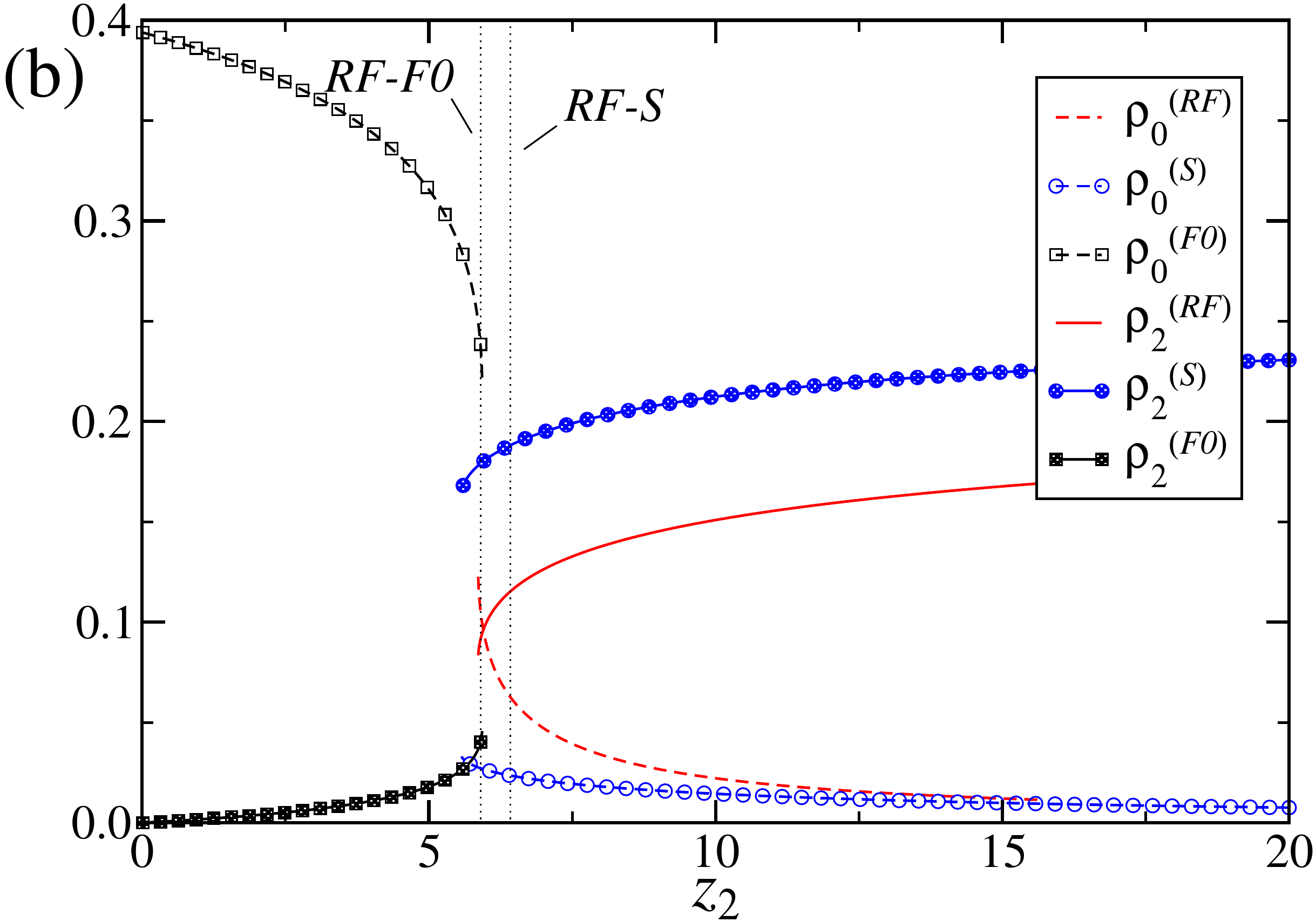}
\includegraphics[width=8.cm]{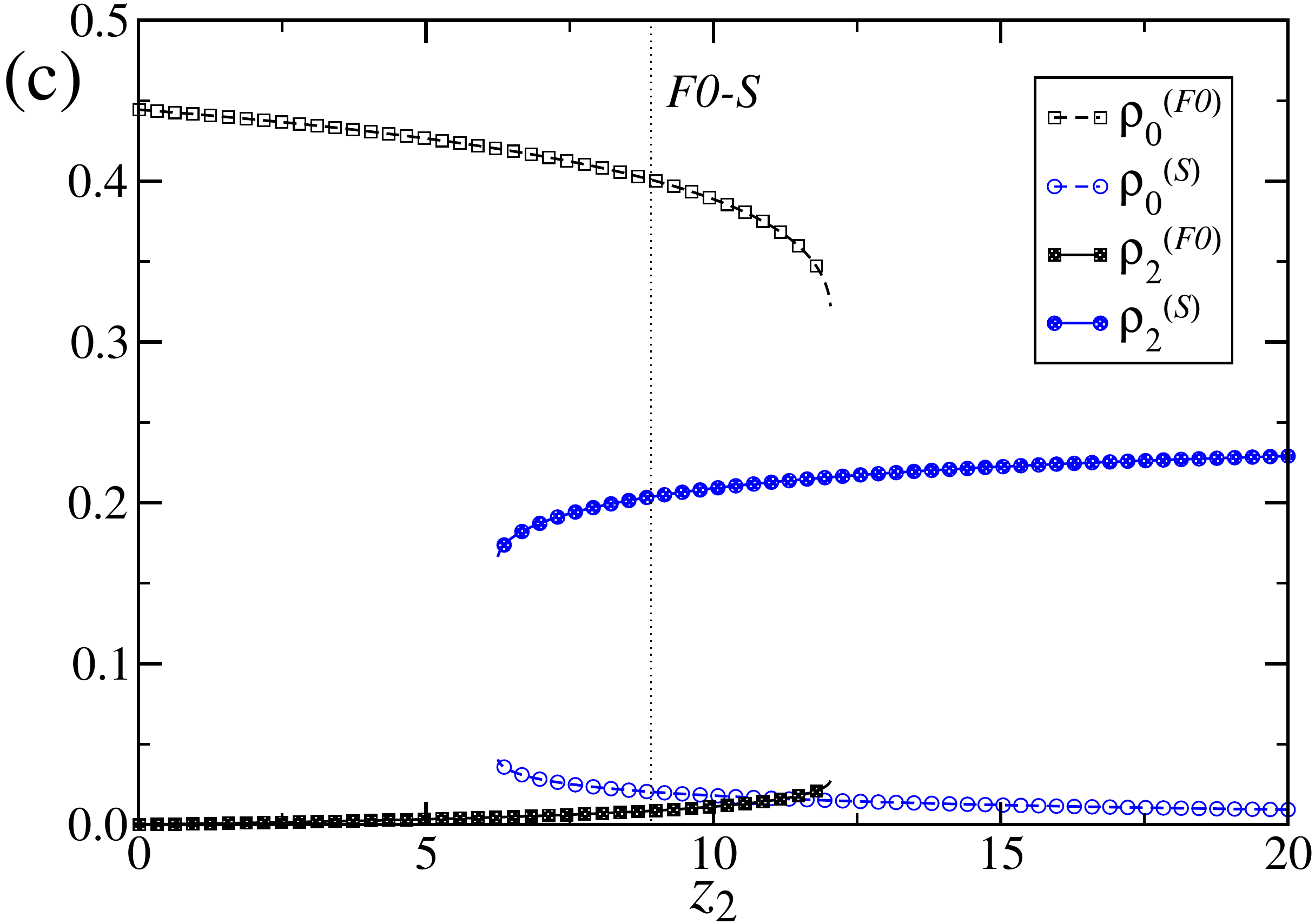}
\caption{Densities of small ($\rho_0$) and large ($\rho_2$) molecules as function of $z_2$ for (a) $z_0=0.5$, (b) $z_0=0.65$ and (c) $z_0=0.8$. The vertical lines are located at the transition points between the phases indicated in each case.}
\label{fig3}
\end{figure}

With exception to the CP and some points where they cross each other, the spinodals do not coincide elsewhere. This implies that there is no continuous transition lines in the phase diagram of the model, and only discontinuous transitions separate the phases. Indeed, we have found three coexistence lines: $RF$-$S$, $RF$-$F0$ and $F0$-$S$, which meet at a triple point (TP) located at $(z_{0,TP},z_{2,TP}) = (0.6774,6.5671)$. See Fig. \ref{fig2}b. The $RF$-$S$ transition line starts at $z_0=0$ (and $z_2=5.7932$ as discussed in the previous subsection) and ends at the TP. It is interesting to notice that initially this line decreases with $z_0$, having an initial slope $(d z_2/d z_0)|_{z_0 \rightarrow 0} = -1$. Then it passes through a minimum - located at $z_0 = 0.2523$ and $z_2 = 5.6522$ -, after which it increases towards the TP. A very similar behavior has been found for the fluid-solid transition in the mixture of 0NN and 1NN particles on the square lattice and approximations to it, though in this case such transition is continuous for small $z_0$ \cite{Jim01,tiago11,tiago15}. Anyhow, in all cases the initial decreasing in the transition lines is certainly a consequence of the effectively attractive depletion interaction among the large particles induced by the smaller ones.

The $RF$-$F0$ coexistence line extends from the CP to the TP, giving rise to a fluid-fluid demixing transition. The $F0$-$S$ transition line starts at the TP and extends to $z_0,z_2 \rightarrow \infty$, being associated to a fluid-solid demixing. For large values of $z_0$ and $z_2$ the fixed point of the $S$ phase tends to, e.g, $R_1^{(A)}=R_2^{(A)}=1$ and $R_1^{(K)}=R_2^{(K)}=0$ for $K=B$, $C$ and $D$, whilst for the $F0$ phase one has $R_1^{(K)}=1$ and $R_2^{(K)}=0$ for $K=A$, $B$, $C$ and $D$. Inserting such limiting values into the (expanded) expression for the free energy, it is easy to demonstrate that the $F0$-$S$ coexistence line is given by
\begin{equation}
z_2 = 3 z_0 + 6 z_0^2 + 4 z_0^3 + z_0^4,
\label{eqCoex}
\end{equation}
for large values of $z_0$ and $z_2$. Hence, in the limit $z_0 \rightarrow \infty$ we have $z_2 \approx z_0^4$ at the coexistence. This is indeed expected, since in the full occupancy limit a 2NN particle effectively occupy a volume of four 0NN ones.

\begin{figure}[!t]
\includegraphics[width=8.cm]{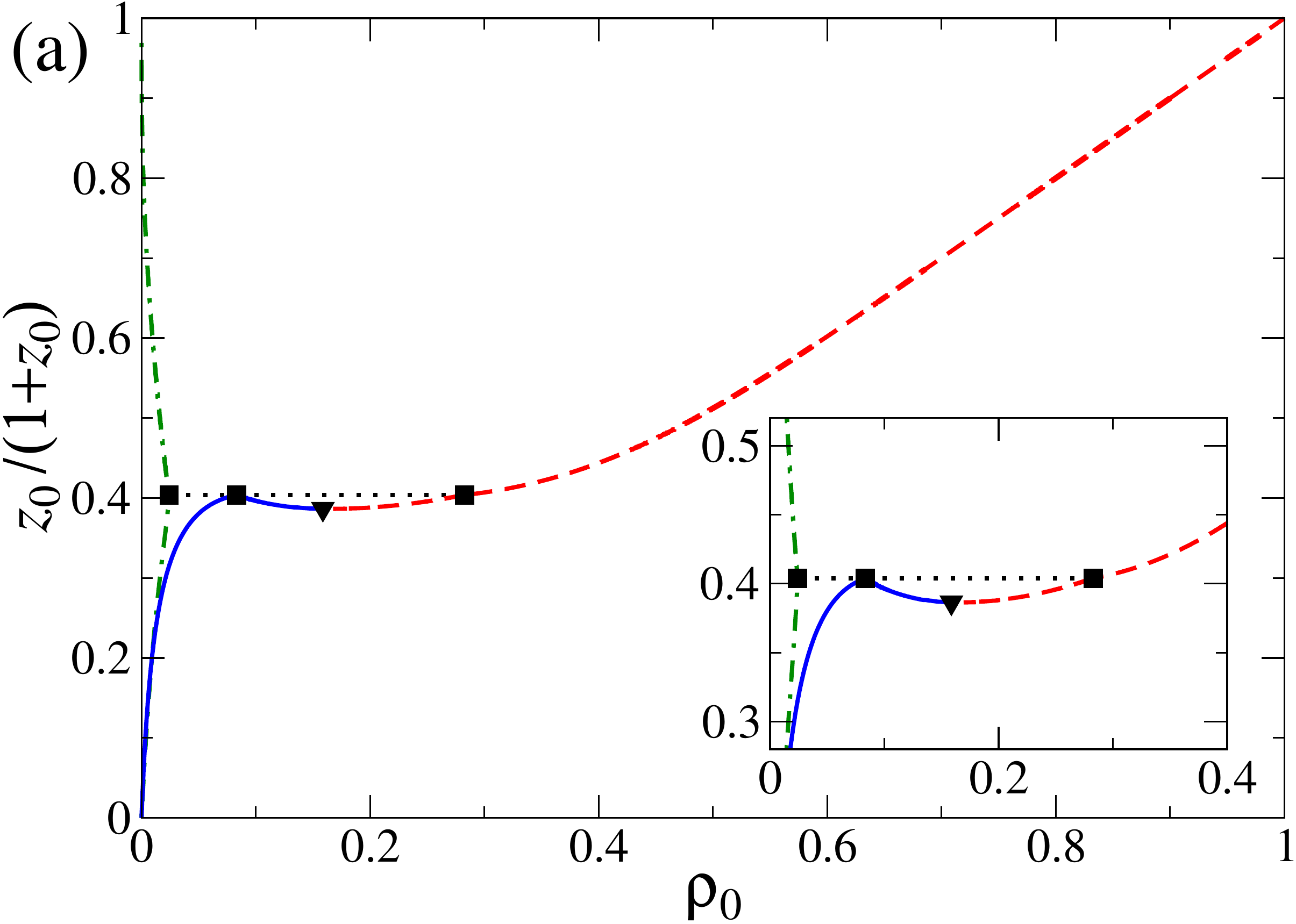}
\includegraphics[width=8.cm]{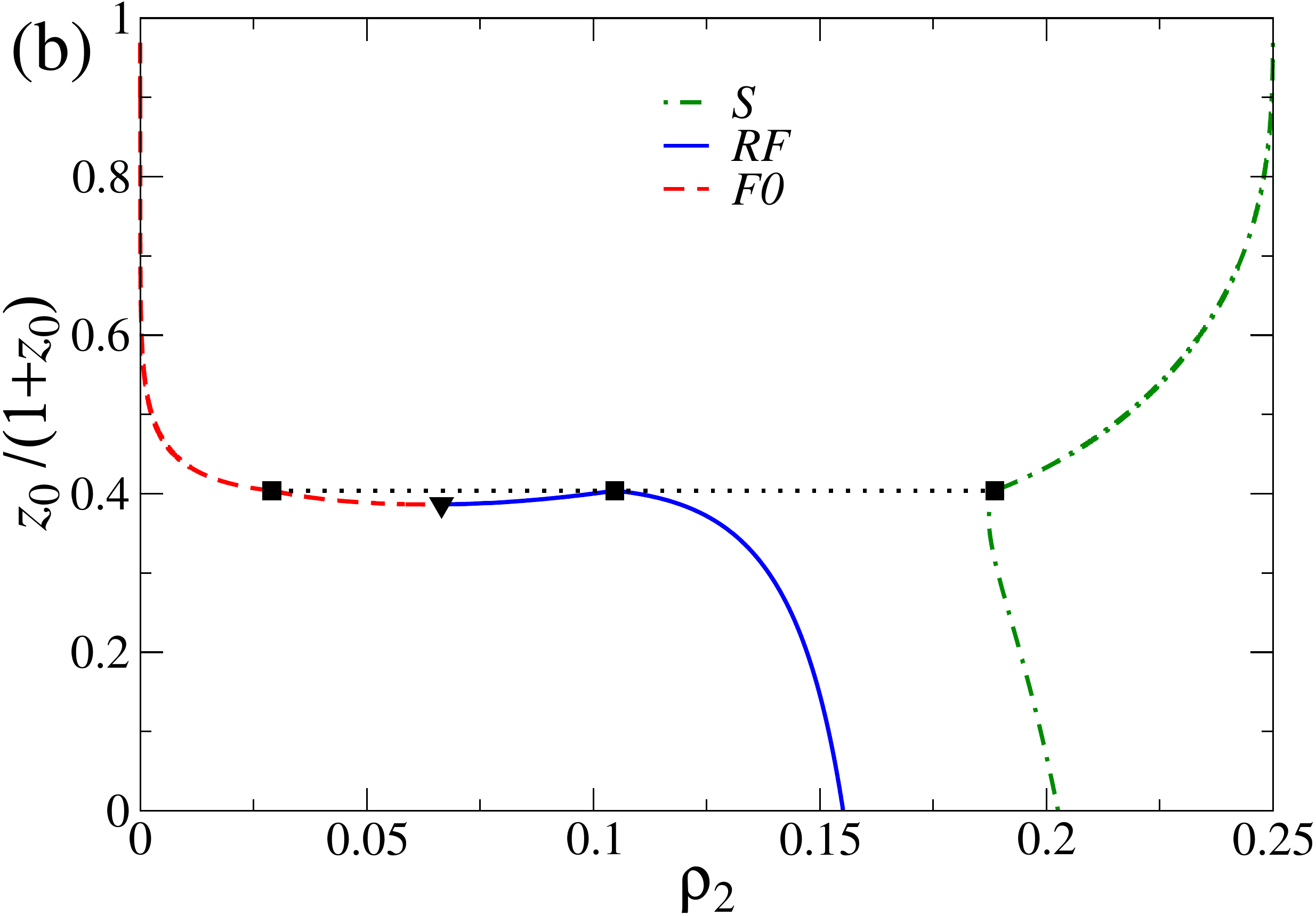}
\includegraphics[width=8.cm]{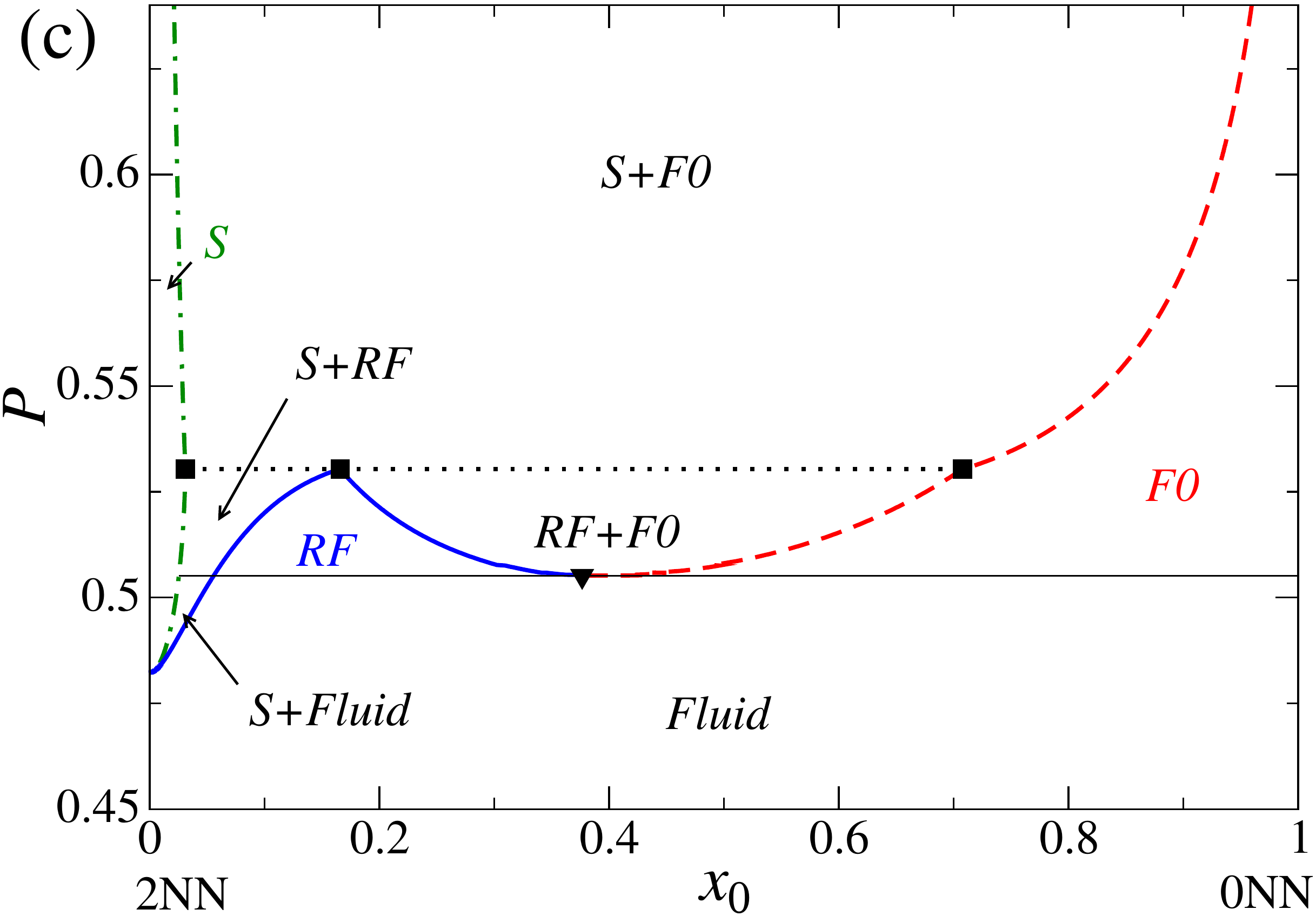}
\caption{Phase diagrams in the variable $z_0/(1+z_0)$ as a function of the densities (a) $\rho_0$ and (b) $\rho_2$; and (c) in the pressure ($P$) versus molar fraction of 0NN particles ($x_0$) plane. The thicker lines are the displayed quantities at coexistence, with exception of the horizontal dotted lines, which indicate the triple point, where the three squares are connected. The critical point is represented by the triangle and the thin horizontal solid line in (c) separates the regions above and below it. The insertion in (a) highlights the region around these points.}
\label{fig4}
\end{figure}

To better understand the differences among the three phases, specially between the two fluid ones, let us take a closer look at the particle densities. The phase diagram in ($\rho_2,\rho_0$) space is presented in Fig. \ref{fig2}c, where one sees that $F0$ phase is featured by large $\rho_0$ and small $\rho_2$, the opposite occurring in $S$ phase, while in the $RF$ one both densities assume intermediate values interpolating between the ones for the other two phases. Figures \ref{fig3}a-c display the densities $\rho_0$ and $\rho_2$ as functions of $z_2$ for three values of $z_0$, chosen such that the three coexistence lines are crossed. The first point to notice is that the $S$ phase is indeed always featured by $\rho_2^{(S)} \gg \rho_0^{(S)}$, as expected, while an opposite behavior is observed in the $F0$ one, namely, $\rho_0^{(F0)} \gg \rho_2^{(F0)}$. This confirms that this phase is indeed dominated by 0NN particles, as already anticipated by the fixed point symmetry. In the RF phase, on the other hand, the densities strongly depend on the activities. Anyhow, in general, one has that $\rho_2^{(RF)}> \rho_0^{(RF)}$ and $\rho_2^{(RF)} < \rho_0^{(RF)}$ at the $RF$-$S$ and $RF$-$F0$ coexistence, respectively. This is also evidenced in the diagrams $z_0/(1+z_0) \times \rho_0$ and $z_0/(1+z_0) \times \rho_2$, depicted in Figs. \ref{fig4}a-b, where the densities at the coexistence lines are shown. Once we know the exact asymptotic form of the coexistence line (Eq. \ref{eqCoex}) and the fixed points there for large $z_0$, the densities of small and large particles can be obtained in such limit. This gives $\rho_0^{(F0)} \approx 1-\frac{1}{z_0}$ and $\rho_2^{(F0)} \approx \frac{1}{z_0^9}$ in the $F0$ phase, and $\rho_0^{(S)} \approx \frac{1}{4 z_0^3}$ and $\rho_2^{(S)} \approx \frac{1}{4} - \frac{1}{4 z_0^3}$ in the $S$ one, in fully agreement with the behaviors presented in Figs. \ref{fig4}a-b. Particularly, this explains the asymptotic linear variation observed in $\rho_0^{(F0)}$ in Fig. \ref{fig4}a. We remark that we have refrained from referring to the $RF$ and $F0$ phases as gas and liquid, respectively, because the total density of particles [$\rho_T = \rho_0 + 4 \rho_2$] at coexistence can be larger in the $RF$ phase. For instance, for $z_0=0.65$, whose densities are displayed in Fig. \ref{fig3}b, one finds $\rho_T^{(RF)} \approx 0.47$ and $\rho_T^{(F0)} \approx 0.40$ at the coexistence point. 

\begin{figure}[!t]
\includegraphics[width=8.cm]{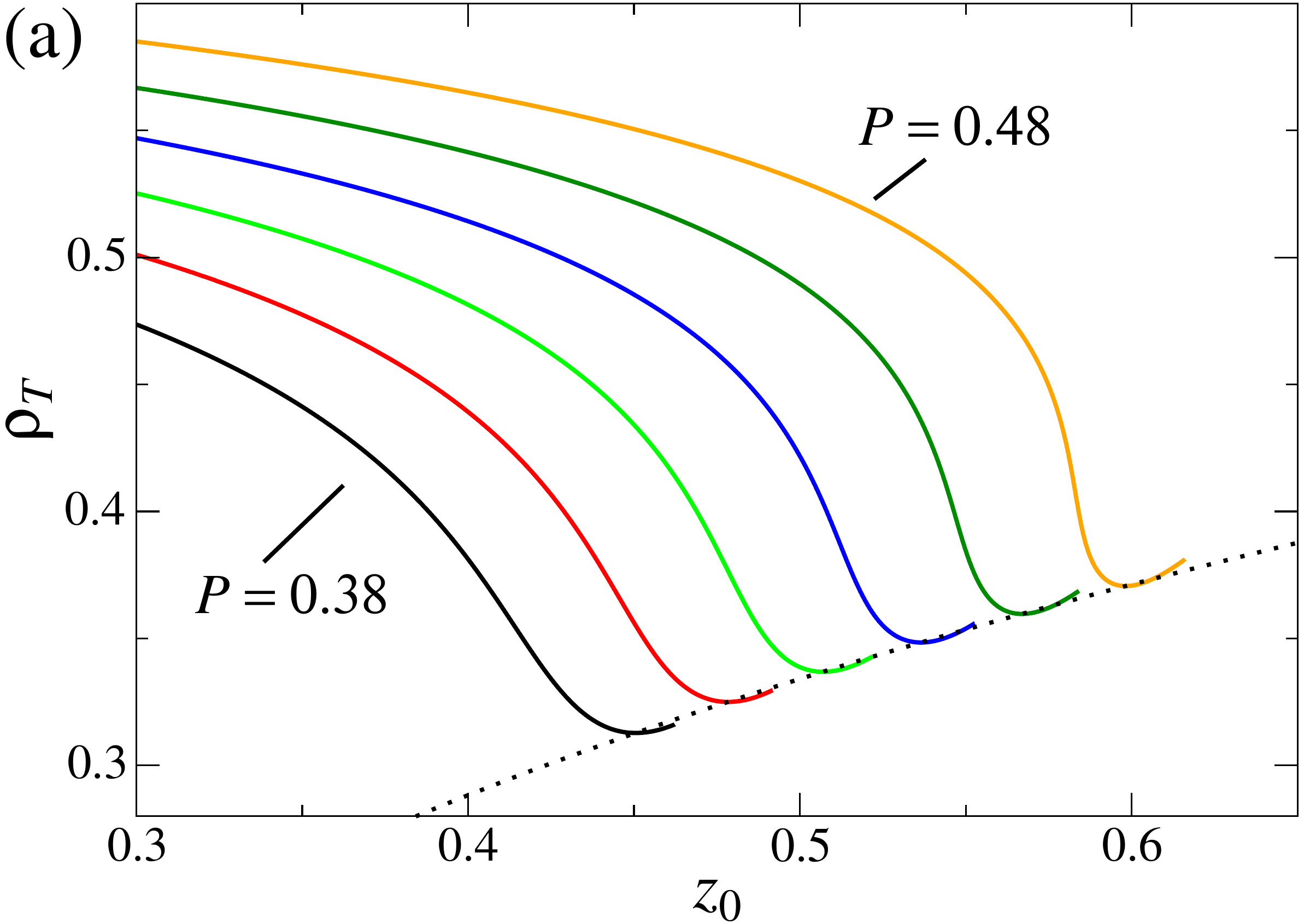}
\includegraphics[width=8.cm]{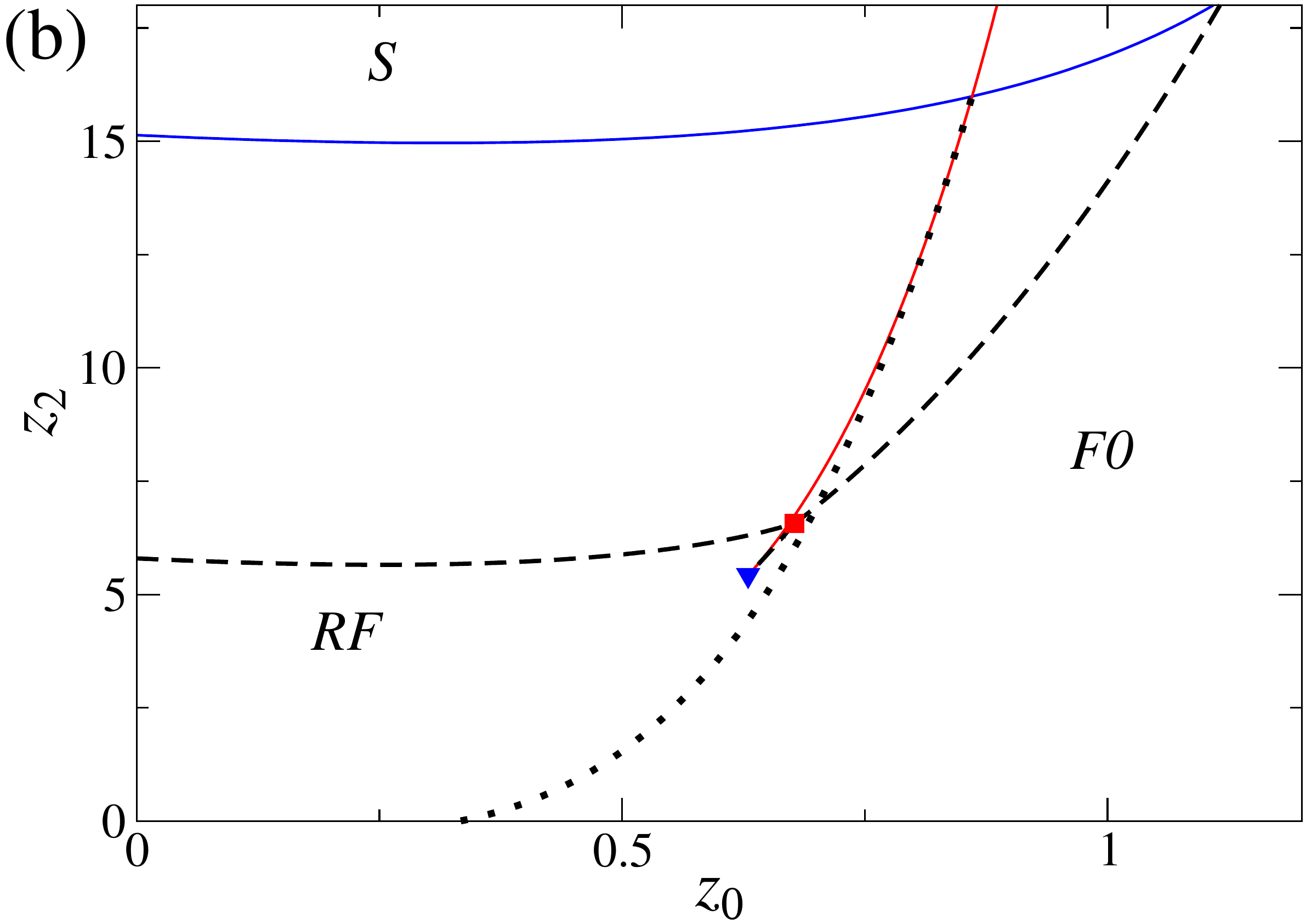}
\caption{(a) Isobaric curves of the total density of particles $\rho_T$ versus $z_0$ for pressures in the interval $[0.38,0.48]$. The dotted line indicate the line of minimum density (LMD). In (b) such curve is plotted along with the ($z_0,z_2$)-phase diagram and some spinodals.}
\label{fig5}
\end{figure}

Figure 4c shows the phase diagram in the pressure-composition ($P,x_0$) plane, where the molar fraction of 0NN particles was defined as $x_0 = \rho_0/\rho_T$. In such diagram, which is qualitatively similar to the $z_0/(1+z_0) \times \rho_0$ shown in Fig. \ref{fig4}a, a detailed description of the several coexistence regions is presented. Althought this is not clear in the scale of Fig. \ref{fig4}c, which highlights the region around the critical and triple points, the (dash-dotted) line for $S$ phase and the (dashed) one for $F0$ phase increase indefinitely as $x_0 \rightarrow 0$ and $x_0 \rightarrow 1$, respectively. In fact, in such limits, which correspond to $z_0,z_2 \rightarrow \infty$ along the coexistence line, it is simple to shown (from the asymptotic behaviors discussed around Eq. \ref{eqCoex}) that $P^{(S)} \sim \ln[1+z_0+z_2] \sim \ln z_0^4$ and $P^{(F0)} \sim \ln[1+z_0] \sim \ln z_0$. So, from the asymptotic behavior of the densities just discussed above, we readily obtain $P^{(S)} \sim -\ln x_0$, as $x_0 \rightarrow 0$, and $P^{(F0)} \sim -\ln[1-x_0]$, for $x_0 \rightarrow 1$.

Interestingly, the total density of particles for the fluid phases, calculated along isobaric curves, display a thermodynamic anomaly characterized by minima in $\rho_T \times z_0$ curves, as shows Fig. \ref{fig5}a. A similar anomalous behavior is also found in $\rho_T \times z_2$ curves. The coordinates ($z_0, z_2$) of such minima for different pressures give rise to a monotonic increasing line of minimum density (LMD), which starts below the CP, at $z_0 \approx 0.333$ when $z_2 \rightarrow 0$, enters the region where the $F0$ phase is stable, crosses the $F0$-$S$ coexistence line, continues within the metastable $F0$ phase and finally ends at a point where the $RF$ and $F0$ spinodals cross each other (see Fig. \ref{fig5}b). This point is located deep inside the region where the solid phase is more stable, so that one may say that the LMD extends to a region where the fluids are super-cooled. We recall that a similar LMD was found in the 0NN-1NN mixture on the square lattice, where it also starts inside the fluid phase, but ends at the tricritical point present in such model, without passing through metastable regions \cite{tiago15}.

\section{Final discussions and conclusions}
\label{secConc}

In summary, we have demonstrated that the athermal binary mixture of 2NN molecules (excluding up to their second-nearest neighbor sites) with 0NN ones (which only exclude the site they are sited on) display a stable fluid-fluid demixing, yielding a very rich thermodynamic behavior with three stable phases in the diagram. The regular fluid ($RF$) and solid ($S$) phases are quite expected, since they are already present in the pure 2NN model. On the other hand, the existence of a second fluid phase ($F0$) featured by a dominance of 0NN particles, although it is somewhat expected because of the nonadditivity of the mixture, turns out to be a very interesting result, in face of its absence in other mixtures of $k$NN particles. We remark that other phases, such as columnar and smectic (observed in some studies of hard cubes with faces parallel to the cubic lattice \cite{Rajeshcubes,LafuenteCuesta2}) are absent in our approach. This is consistent with numerical simulations of the pure 2NN model on the cubic lattice, where only a fluid and a solid phase were found \cite{Panagiotopoulos}. The fluid-fluid demixing observed here, but inexistent in the 0NN-1NN mixtures analyzed (on planar lattices) so far, indicate that in the last case the particle's sizes are not so dissimilar to make this transition stable. Following this though, our results suggest that 0NN-$k$NN mixtures, with $k \geqslant 2$, shall present two stable fluid phases. Moreover, $k$NN-$k'$NN with $k'\geqslant k+2$ can be the condition for fluid-fluid demixing. These points are very important to be analyzed in future works. We anticipate that unfortunately the Husimi lattice built with elementary cubes does not capture the symmetries of the solid phases of $k$NN particles for $k>2$, so that other methods shall be employed to investigate mixtures with larger $k$'s. Particularly important might be Monte Carlo simulations for $k$NN mixtures on the cubic lattice, even to confirm the fluid-fluid demixing observed here, where the differences in particle densities in both fluid phases [as demonstrated in Fig. \ref{fig3}(b)] might be very important to distinguish them and to determine the possible transition lines.

Finally, we notice that the thermodynamic anomalies observed, e.g, in water and other polymorphic fluids usually requires a complex modeling to be reproduced  \cite{polimorfismo}. Our system, similarly to the 0NN-1NN mixture on other lattices \cite{tiago11,tiago15}, demonstrates that a density anomaly can arise in such very simplified models for fluids. Although minima in the isobaric curves of the total density are observed here, whereas in water for example there exists maxima, this indicates that athermal binary mixtures might be useful as a starting building block for the modeling of complex fluids.

\acknowledgments

This work is partially supported by CNPq, CAPES and FAPEMIG (Brazilian agencies).

\appendix
\section{Free energy in a Husimi lattice built with cubes}

A $M$-generation Cayley tree built with cubes has $N_s = 8\times 7^M$ sites at surface and $N_b = 4[7^M-1]/3$ ones in the bulk. Hence, in contrast to regular lattices, the tree is dominated by surface sites in the thermodynamic limit ($M\rightarrow\infty$). Thereby, in order to calculate the free energy in the bulk of such Cayley tree (i.e, in the Husimi lattice), we have to discount the surface contribution. Once the free energy of a $M$-generation tree is $\Phi_M = \tilde{\Phi}_M/k_B T = - \ln Y_M$, following \cite{Gujrati} if we assume that $\phi_b$ and $\phi_s$ denote the free energy per site of bulk and surface sites, respectively, then, we may write also $\Phi_M= N_b \phi_b + N_s \phi_s$. From here, it is a simple matter to see that
\begin{equation}
 \phi_b = \frac{1}{8}(\Phi_{M+1}-7 \Phi_M) = -\frac{1}{8}\ln\left[ \frac{Y_{M+1}}{Y_M^7} \right].
 \label{eqFEap}
\end{equation}
From Eq. \ref{eqY}, we have $Y_M = \left[ A_0^{(M)} B_0^{(M)} C_0^{(M)} D_0^{(M)}\right]^2 y$, while Eq. \ref{eqRRs}a demonstrates that $A_0^{(M+1)}=A_0^{(M)} (B_0^{(M)} C_0^{(M)} D_0^{(M)})^2 R_0^{(A)}$ and similar relations can be obtained for the other sublattices by cyclic permutations of their labels. Substituting these expressions into the rhs of Eq. \ref{eqFEap}, we arrive at Eq. \ref{eqFE}. We note that $y$ and $R_0^{(K)}$, for $K=A,\ldots,D$, do depend only on the ratios and, so, are independent of $M$ in the thermodynamic limit.

\end{document}